\documentclass[11pt,a4paper]{article}
\usepackage{jheppub}
\usepackage{latexsym}
\usepackage{amssymb,amsbsy}
\usepackage{amsfonts}
\usepackage{amsmath}

\usepackage{xcolor}
\usepackage{graphicx} 
\definecolor{brownn}{cmyk}{0,1,1,0.5}
\definecolor{bluen}{rgb}{.1 ,0, .8}

\RequirePackage[T1]{fontenc}
\RequirePackage{times}

\renewcommand{\b}{\beta}

\renewcommand{\d}{\delta}

\newcommand{\ep}{\epsilon}

\newcommand{\la}{\lambda}

\newcommand{\om}{\omega}

\newcommand{\vphi}{\varphi}

\renewcommand{\th}{\theta}

\renewcommand{\exp}{{\rm exp}}

\newcommand{\RR}{{\mathbb R}}

\newcommand{\ZZ}{{\mathbb Z}}

\newcommand{\non}{\nonumber\\}
\newcommand\nn{\nonumber}

\newcommand{\p}{\partial}

\newcommand{\half}{\frac{1}{2}}

\renewcommand{\to}{\ensuremath{\rightarrow\;}}

\newcommand{\refeq}[1]{{(\ref{#1})}}

\newcommand{\twovec}[2]{\left( \!\!
\begin{array}{c} #1\\  #2 \end{array}\!\!\right)}

\newcommand{\twomat}[4]{\left(\!\! \begin{array}{cc} #1&#2\\ 
#3&#4\end{array}\!\! \right)}

\makeatletter
\@addtoreset{equation}{section}

\makeatother

\renewcommand{\ep}{{\boldsymbol{\epsilon}}}
\newcommand{\epi}{{\boldsymbol{\epsilon}}}
\newcommand{\ept}{\tilde{\ep}}
\newcommand{\epti}{\tilde{\ep}}
\let\oldrho\rho\renewcommand*{\rho}{{\boldsymbol{\oldrho}}}
\newcommand{\rhot}{\tilde{\rho}}
\newcommand{\nt}{\tilde{n}}

\newcommand\dsl[1][]{{d\!\! \raisebox{1.4ex}{\_}^{\hspace{1pt}#1}\!}}
\newcommand\du{{d\!\! \raisebox{1.4ex}{\_}u}\,{}}
\newcommand{\dr}[1]{(#1)^{\rm dr}}
\newcommand{\drb}[1]{(#1)^{{\rm dr^*}}}

\newcommand{\Dr}{{\rm\widehat {dr}\,}}
\newcommand{\Drb}{\rm\widehat {dr}{}^*}
\newcommand{\drt}[1]{(#1)^{\rm \widetilde{dr}}}
\newcommand{\veff}{v^{\rm eff}}

\newcommand{\thm}{thermodynamic limit }
\newcommand{\qp}{quasiparticle}

\newcommand{\hp}{h^+}
\newcommand{\tTBA}{torus TBAs{} }

\newcommand{\remark}[1]{}
\title{Low temperature TBA and GHD for simple integrable QFT}
\author{Jacek ~Pawe{\l}czyk}
\affiliation{  Faculty of Physics, University of Warsaw,\\
  Pasteura 5, 02-093 Warsaw, Poland,}
\emailAdd{jacek.pawelczyk@fuw.edu.pl}
\abstract{We derive the low temperature  thermodynamic equations  corrected by virtual processes
for integrable QFT on  large but finite size space circle. 
Obtained TBA's are solved numerically  for the sinh-Gordon model. 
We  also derive corresponding  Euler scale generalized hydrodynamic (GHD) equations
and display them explicitly for small occupation ratio of virtual \qp{s}. The spectrum of velocities for the  linear approximation to GHD is numerically calculated.}
\arxivnumber{2308.05010}
\begin{document}
\maketitle
%
%
%



\section{Introduction}
Effective descriptions of  many-body systems in terms of thermodynamics and hydrodynamics is 
arguably very successful  approach to equilibrium and non-equilibrium phenomena.
Its $(1+1)$ dimensional versions are  useful playgrounds to test our understanding of  complicated phenomena present in  higher dimensions. Thermodynamics of integrable models
is well developed subject \cite{Mussardo:2020rxh}. 
In recent years their hydrodynamical description has been successfully constructed and named generalized hydrodynamics (GHD) \cite{CastroAlvaredo2016,Bertini2016}. 
The approach flourished in numerous developments \cite{Doyon2016,Doyon2017a,Bulchandani2017,Panfil2019,Cubero2019,DeNardis2021,Doyon2022} and extensions 
including effects of diffusion \cite{De_Nardis_2018,De_Nardis_2019} and dispersion\cite{DeNardis2022}. GHD can be applied for quantum systems as well as
classical gasses of solitons \cite{El2021,Bonnemain2022}.
Obtained results were confirmed by numerical simulations (see  also \cite{Moeller2022}) as well as by experiments involving
 one dimensional cold atomic gases \cite{Schemmer2018,Moeller2020,Malvania2020}. For detailed exposition of the subject we refer the reader to  \cite{Doyon2019},\cite{Bulchandani2021} and 
\cite{Essler2023}.

Integrable QFT's  pose new problems  due to virtual processes which come into play. It is known that at finite temperature virtual \qp{s} modify TBA equations \cite{Luescher1986,Luescher1986a,Janik2010a,Bajnok2018}. 
In consequence  they also  change GHD equations \cite{Bajnok2019} but the proposed formulae are hard to deal with: one need to solve generalized BE's which is plagued with technical difficulties.

\begin{figure}[h]\label{fig:torus}
\begin{center}
\includegraphics[width=8cm]{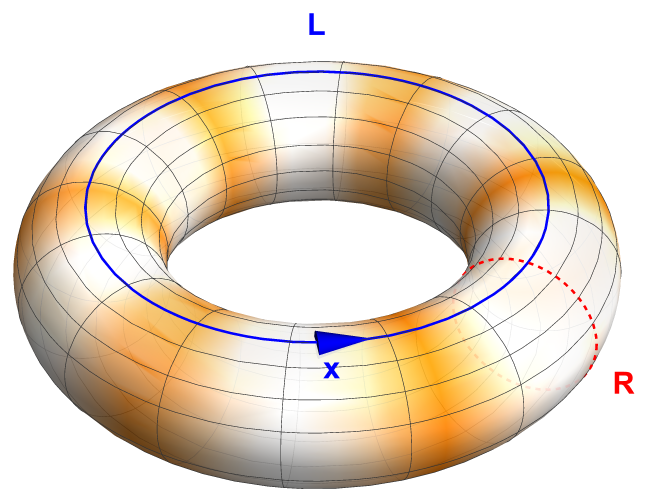}
\caption{The torus as product of two circles: the space size $L$-circle and
$R$ - the temperature  circle. Non-trivial fluid dynamics (imagined by color shades) take place along $L$-circle.}
\end{center}
\end{figure}

Our purpose is to describe thermodynamics and hydrodynamics of an integrable QFT on a space circle of the  size $L$. 
Recall that in QFT the temperature is emulated by going to Euclidean time 
circle of the size  $R$ (see Fig.\ref{fig:torus}). This gives possibility to interchange circles interpretation. Hence one can treat  $1/L$  as the dual temperature for gas of virtual \qp{s} on the circle $R$. We go to \thm on both circles
keeping $R$ and $L$ large but finite and introduce  densities for real (residing on $L$) and virtual (on $R$) \qp{}s. 
Applying duality between the circles we generalize known TBA and GHD equations. The obtained
thermodynamics of real \qp{s} takes into account QFT virtual effects.

The paper is organized as follows. In Sec.2 we recall basic facts about virtual processes in an integrable QFT 
with one type of \qp{}s.   We  take the \thm on the temperature circle and then dualize the equations. The obtained set of  four integral equations   generalizes standard  BYE and TBA equations. Next we
 solve numerically the system for the sinh-Gordon model for some values of free parameters of the theory.
In Sec.\ref{sec:n-ghd} we derive GHD equations employing the same arguments as in Sec.\ref{sec:TBA-gen}.
First we review the construction of the ordinary GHD. Then we write  down the generalized dressing operation which is an essential ingredient of the hydrodynamics derived in the following section.  Next we present some properties of the obtained equations and solve them in linear order in virtual \qp{s} occupation ratio $\nt$. 
In conclusions we summarize our results and propose future research directions.
In Appendix we discuss how the duality reasoning works for  the theory of free fermion field and we include 
 some details of the ShG model, the  dressing operators and 
small $\nt$ expansion of the found GHD equations.
\section{TBA with virtual excitations}
\label{sec:TBA-gen} 
Consider   integrable QFT on large but finite circle of the size $L$.
The temperature of the system is emulated by going to the Euclidean time 
circle of the size  $R$ (see Fig.\ref{fig:torus}).  QFT processes include also virtual
\qp{s} going around that $R$. We want to calculate their contribution going to dual picture which interchanges 
the two circles $L$ and $R$. Hence we can treat  $1/L$  as the dual temperature for gas of virtual \qp{s} on the circle $R$. We go to \thm on both circles
keeping $R$ and $L$ large but finite and introduce continuous densities for real (residing on $L$) and virtual (on $R$) \qp{}s.
In the following we shall  explicitly display formulae  for theories with one type of \qp{}s for which the sinh-Gordon model \cite{ARINSHTEIN1979389,Fring_1993,Koubek_1993} is a benchmark theory. 

Our starting point is the standard TBA equation without virtual \qp s taken into account.\footnote{We decided to denote  $\du\equiv  \frac{d u}{2\pi}$ because this measure appears in almost all formulae.}
\begin{align}\label{TBA0}
\ep_0(\th)= E(\th) - \frac1R\int\!\!\du\Phi'(\th-u )\log(1+e^{-R\ep_0(u)}),
\end{align}
where $\Phi=-i\log S$ is the scattering phase and $E(\th)=m \cosh(\th)$ for mass $m$ \qp{s}. 
Due to QFT phenomena virtual \qp s on $R$ circle are present modifying \eqref{TBA0} \cite{Dorey_1996}.
\begin{align} \label{TBA-N}
\ep_N(\th)=E(\th) + \frac iR\sum_{j=1}^N\Phi(\th-\th_j^+)
- \frac1R\int\!\!\du\Phi'(\th-u )\log(1+e^{-R\ep_N(u)})
\end{align}
Notice that virtual particles rapidities $\th_j^+=\th_j+\frac{i\pi}2$ reside in mirror channel
according to standard nomenclature (see e.g. \cite{Tongeren2016a}). Their
 values are determined  demanding:
\begin{equation}\label{sing}
R\,\ep_N(\th_i^+)=i(2n_i+1)\pi, \quad n_i\in\ZZ
\end{equation}
This yields BE corrected by the ground state contribution (see also \cite{Teschner_2008}).
\remark{BE}
\begin{align}\label{BE-N}
2n_i\pi= & R\,p(\th) +\sum_{j=1}^N\Phi(\th_i-\th_j)
+i\int\!\!\du\Phi'(\th^+_i-u )\log(1+e^{-R\ep_N(u)})
\end{align}
where we have used that the mirror energy is related to momentum: $E(\th^+_i)=i\, p(\th_i)$.  
The two sets of equations \eqref{TBA-N} and \eqref{BE-N} determine 
rapidities of \qp{s} and 
pseudoenergy $\ep_N$, thus the occupation ratio $n=(1+e^{R\ep_N})^{-1}$ of \qp{}s on the space $L$ circle.
 The equations were analyzed to some extent in e.g. \cite{Bajnok2018}.

Going to \thm we assume that the  dominant contribution
to free energy from virtual \qp{}s on the  circle $R$ comes from a density of the latter denoted by $\rhot$. Thus we write \remark{partition} 
\begin{align}
\label{TBA-R1}
\epi(\th)=&E(\th)+ i\!\int\! du\ \Phi(\th-u^+)\rhot(u)
-  \frac1R\!\int\!\!\du\Phi'(\th-u )\log(1+e^{-R\epi(u)})
\end{align}
For $L\to \infty$ we should  have $\rhot=0$ i.e. we should recover standard TBA, which together with
BYE yields density of \qp{s} on the big circle $L$. 
For $\rhot\neq 0$ the BYE gets modified according to \eqref{BE-N}. Differentiating the latter over $\th$  we get:
\begin{equation}\label{BYE-R1}
2\pi i \rhot_t(\th)= E'(\th^+) +i \int\! du\, \Phi'(\th-u)\rhot(u)+ \int\!\!\du\Phi'(\th^+-u )n(u)\epi'(\th)
\end{equation}
where  $\epi'(\th^+)=2\pi i \rhot_t(\th)$ and $\rhot_t$ is total density of  the virtual states.

The  above procedure yields two equations \eqref{TBA-R1} and \eqref{BYE-R1} on three unknown functions $\ep,\ \rhot_t,\ \rhot$. Hence we need to supplement the system by an extra input.
For  $L$ large enough we can use the dual picture of the situation in which we swap the role of $R$ and $L$ i.e.  we can threat $R$ circle as space in which the density  $\rhot$ is determined by the temperature $1/L$ \footnote{See  App.\ref{app:dual} for the discussion of the same approach applied for the theory of free massive fermions.}.
  The equations governing  $\rhot$ are dual cousins of Eqs.  \eqref{TBA-R1} and \eqref{BYE-R1}.
Notice that in the dual picture below we choose $\ept'(\th^-)=- 2\pi i \rho_t(\th)$ instead of 
 $\ep'(\th^+)=2\pi i \rhot_t(\th)$ we had previously. The difference is irrelevant  here due to 
$\Phi(\th^-\!-u )=-\Phi(\th^+\! -u )$ but we have found it useful for the construction of dressing to be discussed in the next section.

 In this way we can write the complete set of equations, hereafter called \tTBA
, on the unknowns: 
$\ept,\ \rhot_t,\ \ep,\ \rho_t$. 
\begin{align}\label{TBA-R}
R\epi(\th)=\,& RE(\th)+iR\!\int\! \du\ \Phi(\th-u^+)\nt(u)\rhot_t(u)
-  \!\int\!\!\du\Phi'(\th-u )\log(1+e^{-R\epi(u)})\\
\label{BYE-R}
R \rhot_t(\th)=&-i RE'(\th^+) +R\!\int\!\!\du \Phi'(\th-u)\nt(u) \rhot_t(u)
+  i\!\int\!\!\du\Phi''(\th^+\!-u )\log(1+e^{-R\epi(u)})\\
\label{TBA-L}
L\ept(\th)=\,& LE(\th) +iL\!\int\! \du\ \Phi(\th-u^+)n(u)\rho_t(u)
-  \!\int\!\!\du \Phi'(\th-u)\log(1+e^{-L\epti(u)})\\
\label{BYE-L}
L\rho_t(\th)=\,&i  LE'(\th^-) +L\!\int\! \du\ \Phi'(\th-u)n(u)\rho_t(u)
- i\!\int\!\!\du\Phi''(\th^--u )\log(1+e^{-L\ept(u)})
\end{align}
where $n(u)=(1+e^{R\epi(u)})^{-1},\ \nt(u)=(1+e^{L\epti(u)})^{-1}$ and we have rescaled $2\pi \rhot_t\to\! \rhot_t$, $2\pi \rho_t\to\! \rho_t$ in order to  use the  integration measure $\du$ everywhere. In the limit $L\to\infty$ the occupation ratio $\nt\to 0$ thus the  virtual particles contribution
to $\ep$ vanishes  so \eqref{TBA-R} and  \eqref{BYE-L} become standard TBA and  BYE, respectively.  

It is well known that integrable systems equilibrate to GGE \cite{2007PhRvL..98e0405R,Mossel2012,2016JSMTE..06.4007V}. 
Thus, in general, inhomogeneous term of  \eqref{TBA-R} should be replaced: 
\begin{align}
R E(\th)\to w(\th)=\sum_\la \b_\la h_\la(\th),
\end{align}
where $h_\la$'s form a complete set of functions on the rapidity space and $\b_\la$ some constants characterizing the GGE state. As the first two (parity invariant) functions we can take $h_0=1,\ h_2=\cosh(\th)$. Thus e.g.  $\b_0=\mu,\ \b_2=R\, m$ are (rescaled by $R$)
chemical potential and mass, respectively.
Notice that \refeq{BYE-R} is not modified in GGE because BYE is just a quantization condition.
On the other hand we expect that the TBA equation for density of \qp s on $R$ at temperature $1/L$
\refeq{TBA-L} stays untouched because there is no dynamics along $R$ circle. In other words, we can not manipulate virtual quasiparticles as we do with physical \qp{s}.
\remark{potentials}
Likewise there is no change in to \eqref{BYE-L}.

We can also go to dimensionless quantities $R\rhot\to \rho,\ R\ep\to  \ep,\ L\rho\to \rho,\ L\ept\to  \ep$.
Then the equations depends only on dimensionless products $R m$ and $L m$.
Of course the occupation ratios $n,\ \nt$ depends only on the latter. Thus non-negligible $n,\ \nt$ 
requires $R m$ and $L m$ to be quite small what says that \qp{s} mass should be very small $\sim 1/R,\ 1/L$. 
The average number of \qp{s} in the system as expressed by dimensionless densities is given by:
$N=\int\du n(u) \rho_t(u),\ {\tilde N}=\int\du \nt(u) \rhot_t(u)$. 
The same remark holds for GHD which depends only on 
occupation ratios.

\remark{tTBA}

\subsection{Numerical solution of the TBA equations for the sinh-Gordon model.}
\label{sec:solve}
The equations (\ref{TBA-R}-\ref{BYE-L}) can be solved numerically. We did it for the sinh-Gordon (ShG) model which serves as a benchmark integrable QFT.  
 It has just one type of \qp{}s \cite{Zamolodchikov_2006,Teschner_2008,Mussardo:2020rxh,Konik2020}.
The expressions for scattering matrices are provided in App.\ref{app:phi}.
\begin{figure}[h]\label{fig:sol}
 \begin{center}
 \includegraphics[scale=.65]{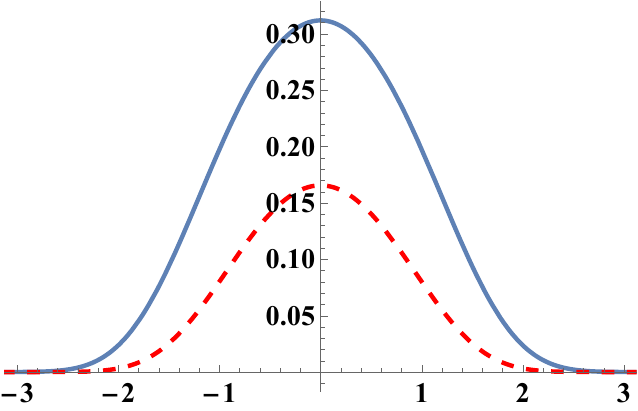}
\hspace{.5cm}
\includegraphics[scale=.65]{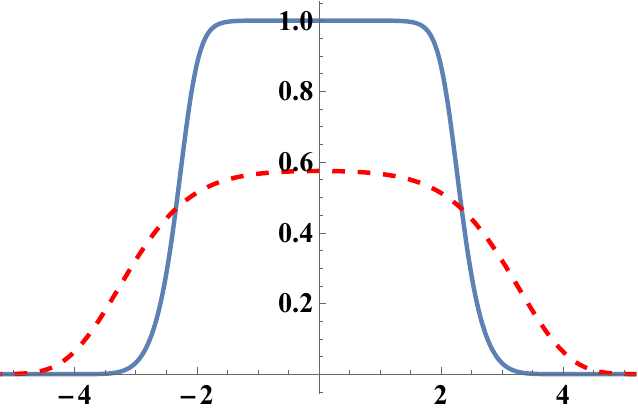}
\end{center}
\caption{Plots of $n(\th)$ (continuum lines) and  $\nt(\th)$ (dashed lines)  obtained from (\ref{TBA-R}-\ref{BYE-L}). The parameters  are (a) $R m= 1,\ L m=1.5,\ a=0.5,\ \mu=0$  (left figure),
and (b) $R m= 0.5,\ L m=0.1,\ a=0.15,\ \mu=-1.5$   (right figure).}
\end{figure}
The results for $n(\th)$ and $\nt(\th)$ are presented on Fig.\ref{fig:sol} for
 two sets of parameters.  One can calculate the average number of \qp{s} in the system given by the formula
$N=\int\du n(u) \rho_t(u),\ {\tilde N}=\int\du \nt(u) \rhot_t(u)$. For the case (a) we get $N\approx 0.32,\ {\tilde N}\approx 0.078$, while for (b), $N\approx 2.3,\ {\tilde N}\approx 2.2$. 
One may wonder if these numbers do not invalidates the whole approach because we expect thermodynamics to be good description for large numbers of particles in the system. The problem 
was addressed in \cite{Malvania2020}. \remark{malv}
The authors claims that experimental data are in agreement with thermodynamics of theoretical integrable models even for few ($N\sim$10) particles. Thus although in the case (a) $N$ looks unreasonably small, the case (b) is much more viable.  It is possible to fine tune parameters of \tTBA to increase 
$N,\ \tilde N$.

\section{Euler scale GHD}\label{sec:n-ghd}
GHD is a very powerful tool devised to provide  description of the macroscopic properties of  integrable models \cite{CastroAlvaredo2016,Bertini2016}. The relevant hydrodynamical quantities are densities and occupation numbers which, besides rapidity, depend on space-time coordinates: $\rho(\th,t,x),\ n(\th,t,x)$, etc., where $x$ is the coordinate along $L$ circle.
Their dynamics is derived from (infinite) current conservation laws of the integrable models under consideration.

\subsection{GHD}\label{sec:ghd}
We start with brief description of technical steps which lead to the standard Euler scale GHD. 
Let  the complete set of functions (enumerated by $\la$) on the space of rapidities  be $h_\la(\th)$. We define space-time densities of charges:
$q_\la=\int d\th\, \rho\, h_\la=\int d\th\,\rho_t\, n\, h_\la$. The dressing of an
arbitrary function $g$ of rapidities is:
\begin{align}
\dr g(\th)=g(\th)+\int \du\, \Phi'(\th-u)\,n(u)\,\dr g(u)
\end{align}
or, in short $\dr g=g+ \vphi\star\,n\dr g$, where $\varphi=\Phi'$ and $\star$ denotes convolution.
The dressing comes into play by BYE 
\begin{align}\label{pdr}
\rho_t=p'+\varphi\star n\rho_t
\end{align}
which can be written as $\rho_t=\dr{p'}$. One can show that (see e.g. \cite{Doyon2019}),
 $q_\la=\int h_\la n\,\dr{p'}=\int \dr{h_\la}n\, p'$ \footnote{Hereafter we shell suppress the rapidity  integration arguments.}.
 Applying the  mirror transformation: 
$q_\la\to i j_\la,\ p\to i E$ we obtain 
$ j_\la=\int E'   n \,  \dr{h_\la}=\int \dr{E'}\, n \, h_\la$.
From there and conservation laws of currents, $\p_t q_\la+\p_x j_\la=0$, one derives the leading order (Euler scale) GHD equations.
\begin{align}\label{ghd}
\dr{p'}\p_t n+\dr{E'}\p_x n=0
\end{align}
The above can be extended to include  diffusive  \cite{De_Nardis_2018,De_Nardis_2019} and dispersive terms \cite{DeNardis2022}.
\subsection{Torus GHD}
Here we shall derive GHD equations relevant for finite size space $L$.  The fluid variable like densities and occupation ratios will depend on time and position $x$ along $L$ but will be uniform in $R$ (temperature) direction. Notice that we need to let quantities related to virtual \qp{s} i.e. $\rhot_t$ and $\nt$ depend on $(t,x)$, too. In fact we shall see that dynamics of $\nt(t,x,\th)$ and  $n(t,x,\th)$ are intertwined by the new 
GHD equations. 

Our procedure will follow the route described in the previous section, but we need to identify the appropriate dressing operation and the charges.
As we shall see both will contain novel ingredients due to the virtual processes of the considered integrable QFT.
\subsubsection{Dressing}
\label{sec:dress}
Dressing  expresses change of a quantity due to interaction with particles of an environment.
It is clear that any change in  the thermodynamic  equations leads to a change of dressing (for   \eqref{TBA-N} and \eqref{BE-N} see \cite{Bajnok2019}). 
The new dressing operator will reflect properties of  all Eqs. (\ref{TBA-R}-\ref{BYE-L}). 

Our starting point is Eq.\eqref{BYE-L} which we write as: 
\remark{ep-rho}
\begin{align}
\label{drp0}
\rho_t=& p'+\vphi\star n\rho_t+\vphi^-\star \nt\rho_t^+\qquad \rho_t^+\in i\RR
\end{align}
where $g^\pm(\th)=g(\th\pm i\pi/2)$. 
\footnote{The above formula seems to be  continuous cousins of that from \cite{Bajnok2019}. Let us mention only that by setting
$\nt=\frac{2\pi i}{\ep'} \d(u-\th^+_i) $ ($\th^+_i$ are as in \eqref{TBA-N}) in \eqref{charge} one recovers charge formula (26) there.
The same holds for the definition of dressing. The relation of \cite{Bajnok2019} to our results is worth careful checking but we shall not dwell on it in this work. }.

The  inhomogeneous linear relations \eqref{drp0} for $\nt=0$ leads to  dressing 
of the standard GHD $\rho_t=\dr{p'}$. For $\nt\neq 0$ the new definition must involve $\rho_t^+$.
Notice that  from \eqref{drp0} we can write:
$\rho_t^+=(p')^++\vphi^+\star n\rho_t+\vphi\star \nt\rho_t^+$
(this  is just derivative of \eqref{TBA-L}). Thus both  equations can be put into vector form,
\begin{align}
\label{tba-L}
\hat \rho_t=&{\hat p'}+\hat\vphi\star \hat n\hat \rho_t,
\end{align}
where 
\begin{equation} \label{ghat}
\hat \rho_t=\twovec{\rho_t}{\rho^+_t},\;
\hat\varphi=
\begin{pmatrix}{\varphi}&{\varphi^-\!}\\{\varphi^+\!}&{\varphi}\end{pmatrix},\;
\hat n=\begin{pmatrix}{n}&0\\0&{\nt}\end{pmatrix}
\end{equation}
The relation \eqref{tba-L} defines the new dressing\footnote{We discuss properties of  the dressing  $\Dr$ in  App. \ref{app:dr}.} $\Dr$:
$\Dr({\hat g})=\hat g+\hat \varphi\,\star \hat n\, \Dr({\hat g})$, 
so it can be rewritten as ($\hat{p'}=(p',i E')$):
\begin{align}
\label{drp}
\hat \rho_t=&\Dr({\hat p'})
\end{align}

Similarly  \eqref{TBA-R} and \eqref{BYE-R} yields ($\hat{E'}=(E',ip')$):
\begin{align}\label{dre}
\hat \ep'=&\Dr({\hat E'})
\end{align}
Of course both \eqref{dre} and \eqref{drp} involve the same dressing operation.

We need to define charges  on the $L$ cycle where the non-trivial dynamics take place. Without virtual processes this is given by the 
BYE from Sec.\ref{sec:ghd} i.e. $\rho_t=p'+\vphi\star \rho$ where $\rho$ is the charge density.
We use this formula as definition of  charge density on $L$. \remark{charge-break}
Introducing $\vphi^{-1}$  we can write
\begin{align}
\label{charge-0}
\rho_t=& p'+\vphi\star (n\rho_t+(\vphi^{-1}\star \vphi^-)\star \nt\rho_t^+)
\end{align}
Thus, in our case,  the charge density is $n\rho_t+(\vphi^{-1}\star \vphi^-)\star \nt\rho_t^+$. \remark{Gauss law}
It consists of two terms: the first one is standard while the second reflects charge renormalization  by virtual \qp{s}.
Hence:
\begin{align}\label{charge}
q_\la=\int h_\la (n\rho_t+(\vphi^{-1}\star \vphi^-)\star \nt\rho_t^+)=\int (h_\la n\rho_t  +\hp_\la \tilde n \rho_t^+)
\end{align}
where $\hp_\la(\th)=\int \dsl u\, \dsl u'\,  h_\la(u)h_\la(u-u')^{-1}\vphi^-(u'-\th)\equiv 
h_\la\star \vphi_\la^{-1}\star \vphi_\la^-$
(see App.\ref{app:phi}). \remark{func} \remark{xp}
The above we shall rewrite as a scalar product between hatted quantities
\begin{align}\label{s-prod}
q_\la=\int  \hat h_\la\hat n\,\hat\rho_t 
\end{align}
where $\hat h_\la=(h_\la,\ \hp_\la)$.
Notice that \eqref{s-prod} is given by a scalar product for which $\Dr$ is a symmetric operator (see also App.\ref{app:op}): 
\begin{align}\label{dr-inv}
\int\Dr({\hat g})\,\hat n\, \hat h&=\int \hat g\,\hat n\, \Dr({\hat h})
\end{align}
Thus we have
\begin{align}\label{charge1}
q_\la=&\int  \hat p' \hat n\, \Dr({\hat h_\la})
\end{align}

\subsubsection{Hydrodynamics}
\label{sec:hydro}
Our construction of the hydrodynamics duplicates what has been described in the beginning of this section 
with some tweaking due to the new dressing we have just defined. 
For any space-time derivative $\p$ we have (see App.\ref{app:dr}):
$\p(\hat n\,\Dr{\hat g})=\Drb\, \p\hat n\,\Dr{\hat g}$ for any $\hat g(\th)$, where $\Drb=(1-\hat n\,\hat\varphi\star)^{-1}$.
Proceeding as for ordinary GHD the current conservation law yields:
$\int \hat h_\la\Drb(\p_t\hat n\,\Dr{\hat{p'}}+\p_x\hat n\,\Dr{\hat{E'}})=0$
which must hold for all $\la$.  Factoring out $h_\la$ we get
$\int h_\la\,(\hat1\cdot\Drb (\p_t\hat n\,\Dr{\hat{p'}}$ $+\p_x\hat n\,\Dr{\hat{E'}}))=0.
$
where $\hat1=(1,\vphi^{-1}\star \vphi^-\star)$. 
Because the the set of functions $h_\la$ is complete on the rapidity space we can 
finally write the final form of the new Euler scale GHD equations.
\begin{align}\label{GHD-torus}
\hat1\cdot\Drb{(\p_t\hat n\,\Dr{\hat{p'}}+\p_x\hat n\,\Dr{\hat{E'}})}=0.
\end{align}

The equations \eqref{GHD-torus} intertwine dynamics of two occupation ratios: the physical ($n$) and virtual ($\nt$) quasiparticles, hence contrary to the standard GHD, they do not form closed system.
We need an extra relation binding $n$ and $\nt$. This is provided by the TBA equations of Sec. \ref{sec:TBA-gen}. The latter holds because hydrodynamics assumes local thermodynamical (GGE) equilibrium
i.e. space-time dependent versions of \eqref{TBA-R} to \eqref{BYE-L} must be respected. It is known that integrable systems equilibrate to GGE (see also Sec.\ref{sec:TBA-gen})
thus \eqref{TBA-R} must be modified by replacing
$R E(\th)$ by generalized potential $w(\th,t,x)=\sum \b_\la(t,x)h_\la(\th)$. On the other hand 
 we expect that \refeq{TBA-L} and \eqref{BYE-L} stays untouched. These two are enough to relate $n$ and $\nt$ what is necessary to solve torus GHD \eqref{GHD-gen}. One can tackle the problem by numerical calculations or by an approximation e.g. small  $\nt\ll 1$ expansion. The latter will be discussed in the next section.
In summery, the torus TBA's and GHD form a closed set of equations on five functions
of $(t,x,\th)$: $\ep,\ \rhot,\ \ept,\ \rho$ and $w$. 

With the relation $\nt=\nt(\th,n(\th,t,x))$ at hand  one can turn \eqref{GHD-torus} into set if first order differential equations of the form:
$
A(n) \p_t n+B(n) \p_x n=0
$
with $A,\ B$ being some (linear, non-local and non-diagonal in rapidity space) operators depending on $n(\th,t,x)$. We expect that $A$ is invertible  yielding $V=A^{-1}B$ as a "velocity" operator. 
\begin{align}\label{GHD-gen}
 \p_t n+V(n)\, \p_x n=0
\end{align}
The analytic form of  the "velocity" $V$ can not be given,
but  we can state that
contrary to the standard GHD $V$ is non-diagonal in rapidity mixing different modes during evolution. This pose another problem in solving the torus GHD equations.

\subsection{GHD  in $\nt\ll 1$ approximation}

Here we display GHD equations  in small $\nt$ approximation. This should be good enough to capture  all the physics we can expect from the approach of Sec.\ref{sec:TBA-gen}.
For technical details  of the small $\nt$ expansion consult App.\ref{app:nt} where the full expressions up to linear order in $\nt$ but to all orders in $n$ are given.
Here for the sake of brevity  we shall farther cut-off the expansions to the order ${\cal O}(\nt n)$ i.e  we use \eqref{app:GHD-nt} which already illustrates  new phenomena of our GHD. 
As we shall see the equations have structure similar to that of ordinary generalized hydrodynamics of Sec.\ref{sec:ghd} with some new features, notably the effective velocity matrix will not be diagonal in rapidity modes even in the linear approximation.
For $L\to\infty$ we get $\nt\to 0$ thus also $\p\nt\to 0$ what leads to the standard Euler GHD what it clearly visible in the explicit results presented below.

At leading order in $\nt$ keeping terms up to ${\cal O}(\nt\,n )$ the resulting hydrodynamic equations  contains only few terms \footnote{For the discussion of one dimensional hyperbolic conservation laws see \cite{Bressan2023}.}.
The obtained equations we put in the form \eqref{GHD-gen}. It is convenient to write the equations for $n_E(t,x,\th)=E(\th) n(t,x,\th)$ instead.  Thus now $\dot n_E+ V_E \p_x n_E=0$ where
\begin{align}\label{Ve}
V_E&\approx \frac{ \dr {E'} - i\vphi^+\star(\nt\,E)}{ \dr {p'}- i\vphi^+\star(\nt\, p) }
+2\pi\,L\, i(\vphi^{-1}\star\vphi^-)\star\nt\, (E \, i\Phi^+ - p \, i\Phi^+ (p/E))\star
\end{align}
\remark{n0-six}
The first term in $ V_E$ is diagonal and the second  non-diagonal in rapidities. The equation is parity invariant $n(\th,t,x)\to n(-\th,t,-x)$.


\subsubsection{Linear GHD for ShG model}
Here we shall consider the simplest situation of linearized GHD equations (see e.g. \cite{Panfil2019}). Thus $ V_E$  will be calculated for a background uniform in space and time.  This we take
$n_0(z)=(1+e^{(1+\d \tau \sin 2\pi z)\cosh\th+\mu})^{-1}$, 
with $\d \tau =0.3,\ \mu=-0.7$ with some $z$ specified later.

We take $n(t,x,\th)$ to be space-time harmonic mode on the circles $L$
\begin{align}
n(t,x,\th)=e^{i(2\pi k\, x/L-\om\, t)}f_n(\th),\quad k\in\ZZ
\end{align}
Then $f_n(\th)$ are eigenvectors of $ V_E(n_0)$
\newcommand{\vv}{v}
\begin{align}\label{eigen-v}
 V_E(n_0)f_\vv=\vv f_{\vv}
\end{align}
where $\vv=\frac{\om L}{2\pi k}$. We found that the spectrum $\vv$ is parity invariant i.e. for each $\vv$ there is one $-\vv$ mode. This reflects parity property of \eqref{Ve}.
For $\nt=0$,  $ V_E$  is the diagonal  effective velocity of the ordinary Euler scale GHD,
$v_0^{eff}=\frac{{\rm dr}({E'})}{{\rm dr}({p'})}$.  For large rapidities $|\th|\to \infty$, both $v_0^{eff}$ and  $\vv$ tend to $E'/p'$. 

\begin{figure}\label{fig:dv}
\begin{center}
\includegraphics[width=7cm]{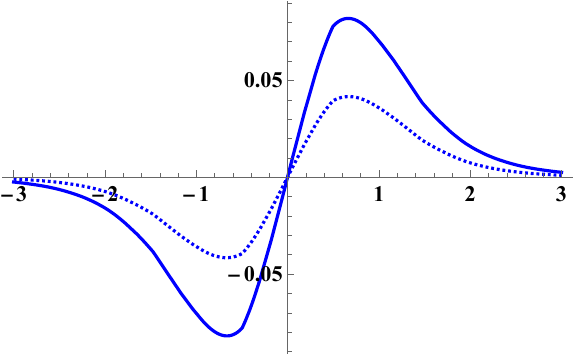}
\hspace{.5cm}
\includegraphics[width=7cm]{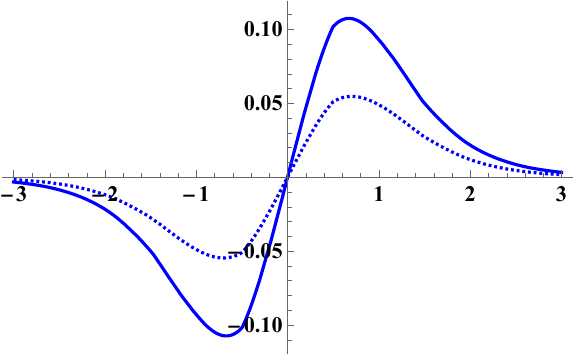}
\end{center}
\caption{The spectrum $\vv$ of $V_E$ compared with group velocity $E'/p'$ calculated for $n_0(z)=(1+e^{(1+\d \tau \sin 2\pi z)\cosh\th+\mu})^{-1}$, 
with $\d \tau =0.3,\ \mu=-0.7$. Figures show  $\vv-E'/p'$ (continuous lines) and  $\veff_0-E'/p'$ (dotted lines) for $z=1/4$ (left figure) and $z=3/4$ (right figure).
The other parameters are:  $Rm=1,\ Lm=1.5,\ a=-0.5$.}
\end{figure}

We solved the eigenproblem \eqref{eigen-v} numerically discretizing rapidity space.  Numerics shows that eigenvalues values $\vv$ 
can be monotonically ordered so we did it with the discrete  $\th$'s. Finally we smooth out the 
resulting function $\vv(\th)$ and compare it with $v_0^{eff}$ and $E'/p'$.
The results are presented on Fig.\ref{fig:dv}.  One can see that $\nt$ corrects the effective velocity roughly by  50\% for $Lm=1.5$. It looks that for higher values of $Lm$ the correction would drop as $\sim e^{-Lm}$.

The velocity eigenvalues $\vv$ and eigenstates $f_\vv$ varies with
$z$ parameter. We can give $z$ physical meaning setting $z=x/L$. In this case one can imagine that $n_0(z)$ parameterize the initial space inhomogeneity. Thus we infer that $\vv$ and $f_\vv$ depend on $x$ making the fluid motion harder to visualize and solve numerically.  

\newpage
\section{Conclusions}

We have presented here the low temperature  TBA  and the GHD equations corrected by virtual processes
for integrable QFT on  large but finite size circle $L$. 
The results have been derived using 
duality between the space and the temperature circles. 
The approach can be applied whenever one can use \thm on both circles.   It is worth to mention that  experiments confirmed integrable thermodynamic approach for relatively small probes 
(tens $\mu m$) and surprisingly small number of particles $\sim 10$ \cite{Malvania2020}.
Obtained TBA's have been solved numerically  for the sinh-Gordon model. 
We have also derived Euler scale GHD. The obtained formulae couple dynamics of  two  fluids: the physical  and virtual  quasi-particles. The equations have very intricate structure thus we presented their expansion in the leading order in $\nt$. The obtained velocity operator is non-diagonal and not symmetric in rapidities. The linearized equations can be easily solved yielding spectrum of velocities. 
Numeric solutions for TBA and GHD showed non-negligible corrections due to virtual processes. 

Equations proposed here need further studies.  Our derivation of  \tTBA is based on duality arguments. 
It is of utmost importance to put the claims on more solid ground. 
One could analyze virtual corrections  by other methods e.g. following  \cite{Bajnok2019} and compare with our proposal. This would require the extensive numerical simulations.  The obtained GHD equations have intriguing structure worth of further investigations. 
It would be interesting to extend the present considerations to other relativistic integrable field theories e.g. to the sine-Gordon model \cite{Mussardo:2020rxh}.
It is also worth to study  higher order terms of the hydrodynamical expansion \cite{DeNardis2022}.
We are planning to address these issues in future works.

\acknowledgments
The author thanks Milosz Panfil, Rebekka Koch  and  Balázs Pozsgay for discussions,  Milosz Panfil  for insightful  remarks on the manuscript,  Gabriele Perfetto for nice communication.

\appendix
\section{Appendix}
\label{app:op}
The appendix contains some extra arguments and technicalities relevant for the body of the paper. We start presenting 
the duality argument of Sec.\ref{sec:TBA-gen} applied for the theory of free fermion field. 
Then we describe some basic properties of the ShG integrable model. Then we explain new notation which will be very useful for small $\nt$ expansion presented in App.\ref{app:nt}. 
We also derive some  properties of the dressing operators.

\renewcommand{\dr}{\mbox{dr}\,}
\renewcommand{\drb}{{\rm dr^*}}
\renewcommand{\drt}{\mathbf{\widetilde{dr}}}
\newcommand{\drtb}{\mathbf{\widetilde{\overline{dr}}}}

\subsection{Duality for free fermion}\label{app:dual}
Here we shortly discuss how the duality works 
for non-interaction theory of a massive fermion (the mass has been rescaled into $L,R$).
The exact partition function  $Z$ of the theory is 
\begin{align}\label{app:Z}
\log Z=L\, {\cal E}_0(R)+\sum_{n\in\ZZ}\log(1+e^{-L E_n(R)})
\end{align}
where ${\cal E}_0(R)=\int \frac{dp}{2\pi} \log(1+e^{-R E(p)})$, $E(p)=\sqrt{p^2+1}$ and 
$E_n(R)=\sqrt{(\frac{2\pi n}{R})^2+1}$. ${\cal E}_0(R)$
is the ground state energy of the fermion on the circle $R$ and the sum comes from particles circling around $R$.

One can calculate ${\cal E}_0(R)$  by standard thermodynamic method i.e.  
interpreting $R$ as inverse temperature of gas of  \qp{s} residing on $L$-cycle. 
One can applies here the dual picture in which $L$ is the inverse  temperature of the gas of  \qp{s} residing on $R$-cycle. This approximates
the second term of \eqref{app:Z} by $R\, {\cal E}_0(L)$. The resulting partition function is thus ${\log Z\approx L\, {\cal E}_0(R)+R\, {\cal E}_0(L)}$. 

\subsection{ShG scattering matrices}
\label{app:phi}
ShG is 2d integrable, relativistic QFT model which  has just one type of \qp{}s \cite{Zamolodchikov_2006,Teschner_2008,Mussardo:2020rxh,Konik2020}. Its S-matrix factorizes into  two-particle scattering amplitude given by:
\begin{align}
S(\th)&=\frac{\tanh\half(\th+i\pi x)}{\tanh\half(\th-i\pi x)}=
\frac{\sinh(\th)+ia}{\sinh(\th)-ia}
\end{align}
where $\th$ is rapidity and $a=\sin(\pi x)$. The scattering phase $\Phi=-i\, \log S(\th)$ and its derivatives are:
\begin{align}\label{Phi}
\Phi(\th^+) &=-\Phi(\th^-) =-i\log(\frac{\cosh(\th) - a}{\cosh(\th)+a})\\
\vphi(\th)&= \Phi'(\th)=\frac{2a\,\cosh(\th)}{\sinh^2(\th)+a^2}
\end{align}

\subsection{Notation}\label{app:not}

Throughout this Appendix we shall introduce simplified notation which will enable to abbreviate various formulae. Thus we shall treat scattering phases, their derivatives and dressings ($\Phi^+,\, \vphi,\ \dr,...$) as  operators which act to the right. We shall suppress $\star$ (denoting convolution) following scattering phase and its derivatives. The other variables will be  just functions. Moreover we shall use in-line notation for dressing operators e.g. $\dr g=(1-\vphi n)^{-1}g,\ \drb g=(1-n\vphi)^{-1}g$.
Thus  e.g. 
$(\drb\, n\,\vphi^-\nt\, p')(\th)\equiv(\drb\, n\,\vphi^-\star \nt\, p')(\th)=\drb (n(\th)\int \dsl u\, \vphi^-(\th-u))\, \nt(u)\,p'(u)$, which can be also written in  the form (see App.\ref{app:dr}):
$n(\th)\dr(\vphi^-)\nt\, p'$. Mastering this formal manipulations should be helpful to cope with
the expansion procedure of App.\ref{app:ghd-nt}.

\subsection{Dressing operators}\label{app:dr}
The space of rapidity functions $\hat g$ are defined by
$\hat h=(h(\th),\hp(\th))$, $h\in\RR,\ \hp\in i\RR$. In  accordance with the expression for charges \eqref{charge-0} we can write down explicit formula for $\hp(\th)=\int \dsl u\, \dsl u'\,  h(u)\vphi(u-u')^{-1}\vphi^-(u'-\th)=\int \dsl u\, \dsl u'\,  h(u)\vphi(u-u')^{-1}\vphi(u'-\th^+)$. 
Here we have introduced $\vphi^{-1}$ respecting
\remark{2pi} $(\vphi\vphi^{-1})(\th-u)=\int \dsl u' \vphi(\th-u')\vphi^{-1}(u'-u)=2\pi\d(\th-u)$, so
$\vphi\vphi^{-1} g=g$.
We have $(\hp(\th))^-=h(\th)$ and one can check that:  $p^+=iE,\ E^+=ip$.
\remark{f+}
One can find ($\b=\arcsin  |a|$):
\begin{align*}
(\vphi^{-1}\vphi^-)(\th)=&
\frac{4\pi i\, \sin\frac{\pi \b }{\pi -2 \b } \sinh \frac{\pi \theta }{\pi -2 \b }}
{(\pi -2 \b) \left(\cosh \frac{2 \pi \th}{\pi -2 \b}+\cos \frac{2 \pi \b}{\pi -2 \b}\right)}
\end{align*}

The new dressing operators has been defined by relations \eqref{tba-L} and \eqref{drp}.
Here we shall use in-line notation  i.e. we write
 $\Dr\hat g$, where $\Dr=(1-\hat\vphi\,\hat n)^{-1}$.
Also $\Drb=(1-\hat n\hat\vphi)^{-1}$.
Some useful identities respected by $\Dr,\ \Drb$ are the same as for $\dr=(1-\vphi n)^{-1}$ and $\drb=(1-n\vphi )^{-1}$ appearing in ordinary GHD (see Sec.\ref{sec:ghd}).
One can  show  that :
\begin{align}\label{dr-id}
&(a)&&\hspace*{-3cm}~~\hat n\,\Dr=\Drb\,\hat n\quad\non
&(b)&&\hspace*{-3cm}\;\int \Dr({\hat g})\,\hat n\, \hat h=\int \hat g\,\hat n\, \Dr({\hat h})\\
&(c)&&\hspace*{-3cm}~~\p(\hat n\,\Dr\!)=\Drb\, \p\hat n\,\Dr\nn
\end{align}
where $\p$ denotes any space-time derivative.
The arguments for the above goes as follows:  from  $(1-\hat n\,\hat\vphi)\hat n=\hat n(1-\hat\vphi \,\hat n)$ which is 
$(\Drb)^{-1}\,\hat n=\hat n\Dr{}^{-1}$ we immediately get (a) .
The l.h.s. of (b) 
is $\int\Dr(\hat g)\,\hat n\, (1-\hat\vphi\,\hat n)\Dr(\hat h)$. The operator 
$\hat\vphi$ is symmetric due to $\vphi^-(\th-u)=\vphi(\th-u^+)=\vphi(u^+-\th)=\vphi^+(u-\th)$ thus  $\int\Dr(\hat g)\,\hat n\,\hat\vphi\, \hat n\, \Dr(\hat h)=\int (\hat\vphi\, \hat n\, \Dr(\hat g))\,\hat n\, \Dr(\hat h)$ what directly leads to the r.h.s.
(c) is  the consequence of
$\p \Dr=\Dr \hat\vphi\,  \p\hat n\, \Dr$.
\newpage
\subsection{Small $\nt $ approximation}\label{app:nt}

\subsubsection{$\mathbf{\nt=\nt(\th,n(\th,t,x))}$ for $\mathbf{\nt\ll 1}$}
\label{app:tba-nt}
Hydrodynamics assumes local thermodynamical equilibrium. Thus some  local  versions of
Eqs. (\ref{TBA-R}-\ref{BYE-L}) must hold.  We expect that  \eqref{TBA-L} and \eqref{BYE-L} are unchanged.
These can be solved for $\nt=\nt(\th,n(\th,t,x))$. 
 In what follows we shall explicitly determine this relation  in the case  $\nt\ll 1$.
Both equations in the new notation are:
\begin{align}\label{app:tba-l}
\ept=& -i p^++i\Phi^-n\rho_t- \frac1L\vphi\log(1+e^{-L\epti})\\
\label{app:bye-l}
\rho_t=& p' +\vphi\, n\rho_t- \frac iL\vphi^-{}'\log(1+e^{-L\ept})
\end{align}
For $\nt\ll 1$ we can use the approximations: 
$\nt={e^{-L\ept}}({1+e^{-L\ept}})^{-1}\approx e^{-L\ept}$ and
$\log(1+e^{-L\ept})\approx e^{-L\ept}=\nt$. 
Then from \eqref{app:bye-l} one can eliminate $n \rho_t$ :
\begin{align}\label{app:nrho}
\rho_t\approx& p' +\vphi\, n\rho_t- \frac iL\vphi^-{}'\nt\quad
\to\quad n \rho_t\approx \drb(n (p' - \frac iL\vphi^-{}'\nt))
\end{align}
Next we  approximate $\eqref{app:tba-l}$ by $\ept\approx-i p^++i\Phi^-n\rho_t- \frac 1L\vphi\nt$. By the above we obtain:
\begin{align}\label{nt-approx}
\nt\approx  \exp({(i Lp^+-iL\Phi^-n\rho_t+ \vphi\nt)})
\approx \exp( -L(-ip^++i\Phi^-\drb{\,n\, p'}))
\end{align}
Notice that  $\nt$ is non-local (in rapidity space) and non-polynomial expression in $n$.

Directly from \eqref{nt-approx} and (\ref{dr-id}):
\begin{align}\label{p-nt}
\dot \nt&\approx L\,\nt\, (-i\Phi^-\drb\, \dot n\, \dr E)
\end{align}
where we have used relativistic dispersion relation $p'=E$.

\subsubsection{Hydrodynamics}\label{app:ghd-nt}
Hydrodynamic equations derived in Sec. \ref{sec:hydro} do not have quite involved form. 
Luckily for $\nt\ll 1$ 
one can rewrite them in a way which resembles ordinary GHD.  
The obtained formulae are exact in $n$  and up to  the first order in $\nt$ only. 

We start with approximations for dressings  operators.
\begin{align}
&\Dr=(1-\hat\vphi\hat n)^{-1}\approx 
\twomat{\dr+\dr \vphi^-\nt\,\vphi^+n\,\dr}{\;\dr \vphi^-\nt}{\vphi^+n\dr}{1}\\
&\Drb=(1-\hat n\hat\vphi)^{-1}\approx 
\twomat{\drb+\drb\, n\,\vphi^-\nt\,\vphi^+\drb}{\;\drb\, n\,\vphi^-}{\nt\,\vphi^+\drb}{1}
\end{align}
where operators $\dr$ and $\drb$ act on everything to the right. e.g. $\drb\,\dot n\,\dr  p'\equiv\drb(\dot n\,\dr  p')$.

Applying the above we can easily approximate GHD equations \eqref{GHD-torus}. Explicitly,
time derivative terms are:
\begin{align}
&\hat 1\cdot \Drb\,\dot{\hat n}\,\Dr(\hat p')\\
&~~\approx\drb\,\dot n\,\dr  p'
+\drb\, (n\,\vphi^-\nt\,\vphi^+\drb\,\dot n+{\dot n}\,\dr \vphi^-\nt\,\vphi^+n+n\,\vphi^-{\dot \nt} \,\vphi^+n)\,\dr  p'\non
&~~~~~+\drb\,({\dot n}\,\dr \vphi^-\nt+n\,\vphi^-{\dot \nt})\,p^+{}' 
+\vphi^{-1}\vphi^-({\dot \nt}p^+{}'
+(\nt\,\vphi^+\drb\,{\dot n}+{\dot \nt}\,\vphi^+n\,)\dr  p')\nn
\end{align}
One can obtain  space derivative terms replacing $(\dot n\to \p_x n,\, p\to E)$

It is clear that for $\nt=0$ (i.e. for $L=\infty$) we recover Euler scale GHD of \cite{CastroAlvaredo2016,Bertini2016}.
In general the equations entangles dynamics of  $n$ and $\nt$ fluids. Their dynamics is constraint by thermodynamics, concretely, by Eqs. \eqref{app:tba-l} and  \eqref{app:bye-l}.
Hence the system of equations is closed and in principle can be solved numerically.

Now we can  apply the formula \eqref{nt-approx} to get hydrodynamic equations up to linear order in 
small $\nt$ but exact in $n$. Notice that we distinguish order of approximation for both occupation numbers because $\nt$ is not polynomial in $n$. From
App.\ref{app:tba-nt} it follows that $\dot \nt\sim \nt \dot n$. Thus up to ${\cal O}(n\nt)$ we have:
\begin{align}\label{app:GHD-nt}
0=&\dot n\,\dr  p'+{\dot n}\,\vphi^-\nt\, p^+{}' +\vphi^{-1}\vphi^-
({\dot \nt}p^+{}'+\nt\,\vphi^+{\dot n}\, p')\non
&+\p_x n\,\dr  E'+{\p_x n}\,\vphi^-\nt\, E^+{}' +\vphi^{-1}\vphi^-
({\p_x \nt}E^+{}'+\nt\,\vphi^+{\p_x n}\, E')
\end{align}

\bibliography{TBA-T2}

\end{document}